\documentstyle[aps, preprint, psfig]{revtex}

\begin{document}

\draft

\title{Granular Drag on a Discrete Object:  Shape Effects on Jamming}

\author{I. Albert$^{1}$, J. G. Sample$^1$, A. J. Morss$^2$, S. Rajagopalan$^2$, A.-L. Barab\'asi$^{1}$ and P. 
Schiffer$^{2}$\renewcommand{\thefootnote}{\alph{footnote}}
\footnote{Corresponding author:  schiffer@phys.psu.edu}}

\address{
$^1$ Department of Physics, University of Notre Dame, Notre Dame, IN 46556\\
$^2$ Department of Physics and Materials Research Institute, Pennsylvania State University, University Park, PA  
16802
}
\date{\today}

\maketitle

\begin{abstract}
We study the drag force on discrete objects with circular cross section moving 
slowly through a spherical granular medium.   Variations in the geometry of the dragged object change the drag force only by a small fraction relative to shape effects in fluid drag.  The drag force depends quadratically on the object's diameter as expected.  We do observe, however, a deviation above the expected linear depth dependence, and the magnitude of the deviation is apparently controlled by geometrical factors.
\end{abstract}


The drag force experienced by a solid object moving through a fluid is one of the most basic phenomena of fluid mechanics.  Despite its fundamental nature, the complexity of fluid drag and its strong dependence on the exact geometry of the object, require that it be determined numerically or experimentally in all but a few simple cases \cite{FluidBook}.
An analogous drag force exists when an object is dragged through a
granular medium, although the physical origin of granular drag at low
velocities is quite different. When an object moves slowly through a
granular medium, it is resisted by the so-called jamming of the grains
\cite{MECates,AJLiu} which occurs when an applied stress results in the
frustration of local granular motion. This jamming is manifested by the
formation of internal networks of force chains among the grains \cite{CLiu,BMiller,XJia,SNCopper,LVanel} which
resist the motion and then collapse as the object moves through. 

In analogy to drag in a fluid, a natural question arises as to how the shape of the dragged object affects the
net drag force in a granular medium. The object's shape determines the nature of the local
jamming in front of the object and, in particular, the
strength of the jammed state (i.e. at what stress it will collapse).
Previous studies of granular drag in static dense granular media have focused on vertical cylinders
inserted into the grains \cite{Wieghardt,RAlbert,IAlbert}.  Since the average intergranular stress in front of such extended objects increases 
continuously from zero, those measurements cannot easily probe the local
jamming. In this work, we focus instead on the drag experienced by 
discrete objects with circular cross section normal to the grain flow direction,  with the goal of understanding the effects of geometry on the jamming and the consequent drag force.  We find that the drag has the theoretically expected quadratic dependence on the diameter of the object, but that there is an unexpected non-linearity to the dependence on depth.  The strength of the non-linearity depends on the object's shape, and can be minimized by either streamlining the object or by reducing its length in the direction of motion.

The experimental apparatus has been used previously to study the drag on 
vertical cylinders and its time-dependent fluctuations, and it is described in
detail elsewhere \cite{RAlbert,IAlbert,thesis}. For the present experiments 
we measure the drag on an object with circular cross section 
(diameter $d_{obj} = 25.4mm$ unless noted otherwise) which is attached to the end of a vertical support rod \cite{Cylinder}. The center of the object is inserted to a depth $H$ in a container filled with monodisperse
glass spheres of average diameter $d_{g} = 0.9 mm$ \cite{Jaygo}. The container rotates with
constant angular speed while the support rod is attached to an arm that rotates freely around the rotational axis of the container. The object
and its support rod are stopped by a fixed precision
force cell \cite{ForceCell} which measures the combined effective drag force on 
the rod and object. The drag force experienced by the rod and object,
$F(t)$, is not constant, but has large 
stick-slip fluctuations corresponding to the jamming and collapse of 
the grains opposing the motion \cite{IAlbert}. 

The objects included five different shapes: spheres, disks, cut-spheres, 
teardrops, and cones as shown in Fig. \ref{ObjectShapes}. Note that the drawings
are to scale and all objects have the same circular cross sectional
area relative to the direction of flow.  The objects were made of aluminum or brass and had identically prepared, rough sand-blasted surfaces. The speed of
movement through the medium does not affect the granular drag
force at low velocities \cite{RAlbert} and was kept constant at 0.2mm/s.  Note that at these velocities, the drag process is effectively probing a static medium since the grains settle on a much shorter time scale than that required to stress the jammed state to the point of collapse \cite{RAlbert,IAlbert,thesis}.  We are thus probing drag in the static limit where the force is determined by jamming of the grains.  This situation is  rather different from the two previous studies of drag on discrete objects, in material fluidized by vibrations \cite{Ozik} or at high velocities \cite{Wieghardt}, and the results are indeed qualitatively different.

In order to separate the drag on the object alone from that on the support
rod, we also measured the drag force on just the support rod with no object attached. The drag
on the object was then determined by subtracting the contribution of the
rod from the total force. To
test the validity of this subtraction, i.e. if the presence of the rod
affects the drag force on the object, we measured the drag using rods of
varying diameter from $10$mm - $19$mm with two different objects (25 mm diameter sphere and disc). In Fig.
\ref{RodTest}, the upper line (circles) shows the total force
experienced by the system composed of the rod and the sphere for
increasing rod widths. The lowest line (triangles) shows the force on
the rod alone, with no object attached to it. The force on the sphere (squares)
is taken to be the difference between the previous two. As shown on the
graph, the force on an object calculated this way is independent of the
rod diameter to within our uncertainty ($\pm 3$\%) and thus is apparently not 
affected by the presence of the rod to within that precision (although there remains some possibility of a more subtle interaction, as discussed below). Therefore, we show below the drag force on discrete objects with the force contribution from the support rod already subtracted off.

In the case of a vertical extended object (such as a cylinder)
with diameter $d_c$ inserted to a depth $H$ in a granular bed,
the drag force is described by $\overline{F} = \eta\rho gd_cH^2$ where
$\eta$ characterizes the grain properties (surface friction, packing
fraction, etc.), $\rho$ is the density of the glass beads and $g$ is
gravitational acceleration \cite{RAlbert}. This formula can be derived
from a mean field approximation which assumes that the
resisting force increases linearly with depth in proportion to the ambient pressure \cite{thesis}.
Although the presence of fluctuations in the drag force indicates 
that the mean field picture does not completely describe the physics,  
more sophisticated theoretical treatments
\cite{RAlbert} have also produced the same result for the average force. For a discrete object with 
circular
cross sectional area, either approach would predict
$\overline{F} = \beta\rho gd_{obj}^2H$, where $\beta$ describes the properties
of the granular medium (equivalent to $\eta$), $d_{obj}$ is the diameter of
the circular cross-section of object, and $H$ is the depth of immersion for the 
center of the object. 

We tested this expectation by a careful examination of the drag on a
sphere as a function of the sphere's diameter and depth in the medium, and our results are shown in Fig. \ref{SphereGraph}. As seen
in the figure, the diameter dependence of the drag force is reasonably well 
described as quadratic. The depth dependence, however, shows a distinct non-linearity.  Since this non-linearity could be attributable
to the finite size of the container \cite{IAlbert}, we also performed measurements for a smaller sphere, as shown in the figure.  The drag forces on the two spheres showed the same depth dependence (varying only by a constant of proportionality), however, excluding finite size effects as an explanation for the non-linearity.

In Fig. \ref{ShapeComparison} we compare the drag on the different objects with 
circular cross section, showing the depth dependence to demonstrate
how the differences evolve with depth.
We find that the drag force is much less affected by object geometry than for fluid 
drag, with the biggest measured
difference between the highest (disk) and lowest (teardrop) no
more than 30\% (for fluids the variations can be more than 300\% \cite{FluidBook}). Note that all 
of the objects show non-linear depth dependence similar to that of the spheres in Fig. \ref{SphereGraph}.  Furthermore, the drag appears 
to be nearly shape independent at small depths and then separates for the 
different shapes at larger depths.  These data suggested that we fit the depth 
dependence to the form $\overline{F} = AH + BH^n$ where $n > 1$.  Choosing the 
value  n = 2, we find that we can fit the data rather well \cite{exponent}
as shown by the solid lines in Fig. \ref{ShapeComparison} with fit parameters given in Table \ref{table1}.  

Examining the fit parameters, $A$ and $B$, we observe that the coefficient of the linear term is  almost independent of the object's shape, while the non-linear term results in most 
of the variation between the shapes.  This strongly suggests that the 
non-linearity in the depth dependence is associated with geometrical factors in the 
drag for which the simple theoretical expectations do not account.   Since the  
theoretical expectations do account well for the quadratic depth dependence of 
the drag on a vertical cylinder inserted from the top surface, we are forced to 
conclude that geometrical effects are much more important for discrete than for
extended objects. This conclusion is also supported by our measurements of the 
drag on a full vertical cylinder and one which is bisected along a vertical  
plane normal to the flow direction - which differ by only a few percent 
\cite{IAlbert}.  The relative importance of shape effects on the different sorts 
of objects can perhaps be attributed to the fact that grains must travel around 
all sides of the discrete objects, rather than only on either of the two sides 
of the vertical extended objects, and therefore the finite size of the container and the small curvature associated with rotation may have a larger role.

An alternative explanation for the non-linearity in the depth dependence could be based in the coupled nature of the support rod and object system.  The rod coupled to the object sets up a stress-field within the grains which is necessarily different from that of an object being dragged without the support rod.  Furthermore, when the jammed grains collapse to allow the object to advance, they must collapse all the way to the surface of the grains to allow the rod to advance also.  The stress fields induced by the rod and the object must combine to nucleate the collapse, and that interaction could potentially impact the depth dependence.  Although the data taken for spheres in figure \ref{RodTest} and similar data taken for disks demonstrate that the size of the rod does not affect the net drag of the rod and object together, we cannot completely exclude that a rod-size independent effect on the drag could be inducing the non-linearity.

We now consider the various factors which affect the granular drag force on a discrete object.   The friction between the dragged object and the grains might be expected to contribute to the drag force (since there is no boundary layer as in the fluid case), but we previously demonstrated that such a frictional contribution is negligible in the case of vertical cylinders \cite{IAlbert}.   This is verified for the present case of discrete objects with the data in the inset of Fig. \ref{SphereGraph} where we plot the drag on spheres with coefficients of friction varying by a factor of 2.5 and find that it is the same within our experimental uncertainty ($\pm 3$\%) \cite{FrictionExperiment}.  The independence of surface friction is important since it indicates that failure of the jammed states is not nucleated at the interface between the grains and the dragged objects.  Rather the collapse of the jammed state originates at an intergranular contact point, and can be attributed to the compressive rather than the shear stress induced by the dragged object.   
 
Since surface friction does not contribute to granular drag, we must consider the effects of the shape on the jamming, compression, and
eventual collapse of the grains impeding the objects' motion.  Since obtaining geometrical factors even in fluid drag is largely an empirical process, we can expect only to identify what geometrical factors may increase or decrease the drag on a object.  An obvious geometrical factor is streamlining of the object and the resultant effects on the dispersion of the force chains 
from the surface. An object which is more tapered toward the front, such as the sphere or teardrop, will presumably apply a more concentrated force on the grains at its most forward point.  This concentration of force will then lead to a collapse of the jammed grains at a lower total force than for the disk where the stress is spread out more uniformly over a broad area of grains.  Indeed we do observe that the disk and the cut sphere have higher drag than the sphere or teardrop.   To understand the detailed differences in the stress propagation from the shapes, however, would require careful modeling beyond the scope of the present paper.

A second and unexpected geometrical factor in the drag force is the length of the dragged object in the flow direction.  We observed this effect by measuring the drag on disks of varying length (Fig. \ref{DiskWidth}), and we found that the drag increases linearly with disk length.  Although it may be natural to attribute this increase to friction with the edges of the disks, we found that the force was unchanged for teflon and rough metal disks -- in agreement with the independence of friction discussed above. Thus longer disks somehow create a jammed state which can withstand a larger applied force before collapsing.   This effect could be understood by assuming that the sides help to distribute the force, i.e. larger disks are in contact with more grains and more force chains can emanate from their sides. This would reduce the local stress on the grains in front of the disk, where the stress is maximized, allowing for a larger force to build up before the grains collapse.  Alternatively, the longer disks may impede collapse of the grains behind the object, and therefore may increase the strength of the jammed state in front of the disk.  It should be noted that only the length of the object at maximum diameter seems to affect the drag, since the teardrop has less drag than the sphere despite being longer in the direction of flow.  This effect may explain why the cone (which has a wide section at maximum diameter) has much larger drag than the teardrop despite being similarly tapered at the front.  Detailed  modeling of the jamming and flow of grains around the object will be required to truly understand this effect, but it is notable that the difference in the drag on different width discs has a super-linear depth dependence - consistent with other geometrical effects on the objects' drag.

In summary, we have studied the drag force experienced by discrete objects moving through static dense granular media.  We find that there is an unexpected non-linear component to the drag which depends on the geometry of dragged object.  We identify basic geometrical factors contributing to the drag process, but detailed modeling of local three-dimensional stress propagation in granular media is needed to elucidate the details of this process.

We gratefully acknowledge the support of NASA grant NAG3-2384, the Petroleum Research Fund administered by the ACS, the Alfred P. Sloan Foundation, and
 NSF grants PHYS95-31383, DMR97-01998, and DMR00-97769.

\begin{table}[h]
\begin{center}
\caption{Coefficients for a $\overline{F} =AH + BH^2$ fit of the depth dependence data for different 
objects.}
\begin{tabular}{cccc}

Object & A$\times 10^2 (N/m)$ & B$\times 10^4 (N/m^2)$\\
\hline\hline
sphere & $3.33\pm 0.03$ & $2.8 \pm 0.3$\\
\hline
disk & $3.23\pm 0.03$ & $3.7 \pm 0.3$\\
\hline
cone & $3.36\pm 0.02$ & $3.0 \pm 0.2$\\
\hline
teardrop & $3.34\pm 0.04$ & $2.6 \pm 0.3$\\

\label{table1}
\end{tabular}
\end{center}
\end{table}

\begin{figure}
\caption{Schematic representation of the objects dragged in the medium. 
The drawings are to scale and the objects have the same circular cross-sections 
in the direction of flow.  The angle $\Theta$ is $90^{\circ}$ and the cut sphere 
has $a=15$ mm.}
\label{ObjectShapes}
\end{figure}
\begin{figure}
\caption{The effect of the support rod on the drag force on a 
sphere with $d_{obj}=25.4$ mm at $H=100$ mm depth.  We take the drag on the sphere to be the difference between the total drag (on the sphere and the support rod) and the drag on the rod alone.  This quantity is constant for spheres and also for disks (not shown), demonstrating the validity of this analysis.}
\label{RodTest}
\end{figure}
\begin{figure}
\caption{The drag force on a sphere. 
a) A log-log plot of the dependence on the diameter of the sphere for depth  $H=100$ mm.  The solid line has slope of two, demonstrating a quadratic diameter dependence.  b)A log-log plot of the depth dependence for two spheres of different diameters $d=25.4$ mm (triangles) and $14.2$ mm 
(circles). The straight lines have slope 
of 1.0, demonstrating the non-linearity of the depth dependence.  The inset shows the depth dependence of the drag for a rough metal sphere (squares) and a smooth teflon sphere (triangles) plotted on a linear scale.  These data demonstrate that the surface friction of the objects does not affect the drag to within the precision of our measurements.}
\label{SphereGraph}
\end{figure}
\begin{figure}
\caption{Comparison of the depth dependence of drag force on different shapes (with the drag on the support rod subtracted off).  The solid lines are fits to the data as described in the text.}
\label{ShapeComparison}
\end{figure}

\begin{figure}
\caption{The drag force on disks as a function of length in the direction of flow (with the drag on the support rod subtracted off).  Note that the increase in drag with increasing disk length is not due to surface friction since the data were unchanged when teflon disks were substituted for the rough metal disks.}
\label{DiskWidth}
\end{figure}


\begin{references}
\bibitem{Cylinder}{
The diameter of the cylinder ($d_c=10$ mm) was chosen to minimize the pertubation 
on the attached object's drag, but be sufficiently thick to be rigid. The object 
is attached on the bottom to a 12 mm long and 6 mm wide cylindrical extension.
}
\bibitem{Jaygo}{
Unless otherwise noted all experiments have been carried out with glass spheres 
of average 
diameter of 0.9 mm from Jaygo Inc., 675 Rahway Avenue, Union, NJ 07083.
}
\bibitem{exponent}{The data could be fit well for a range of the exponent $n = 
2-3$.}
\bibitem{FrictionExperiment}{
We measured the friction angle $\varphi$ for the rods sliding on a plate covered
with glass beads glued onto its surface.($\mu=\tan(\varphi)$).
}
\end{references}
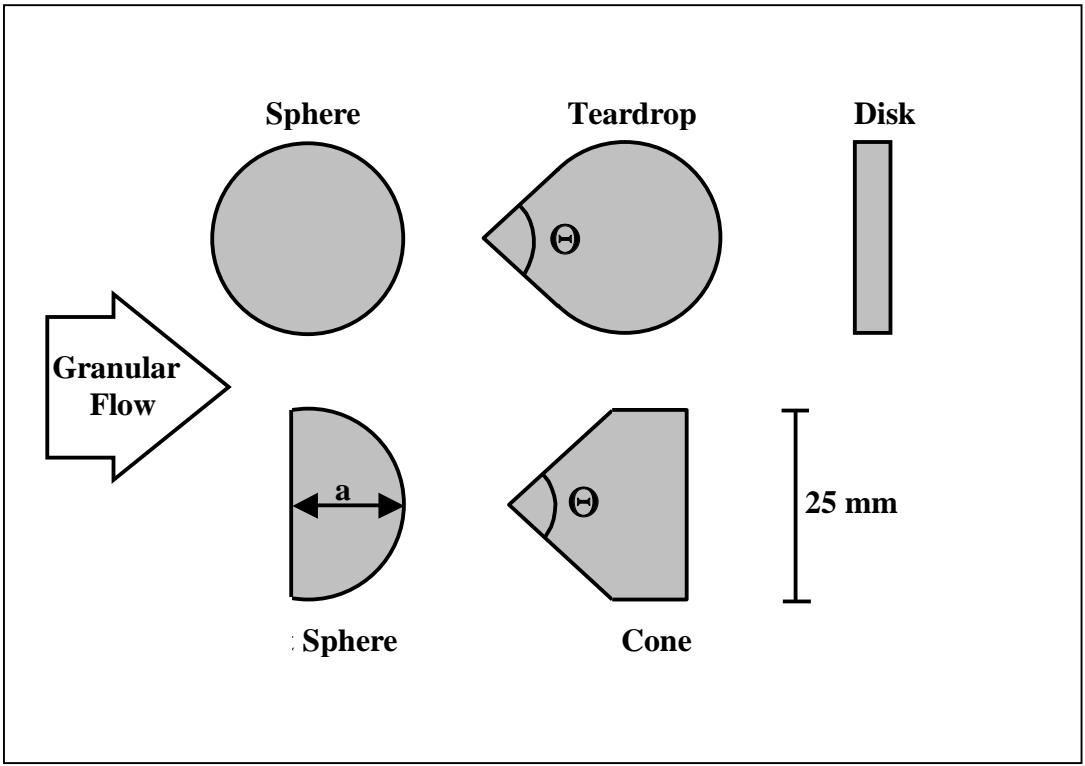
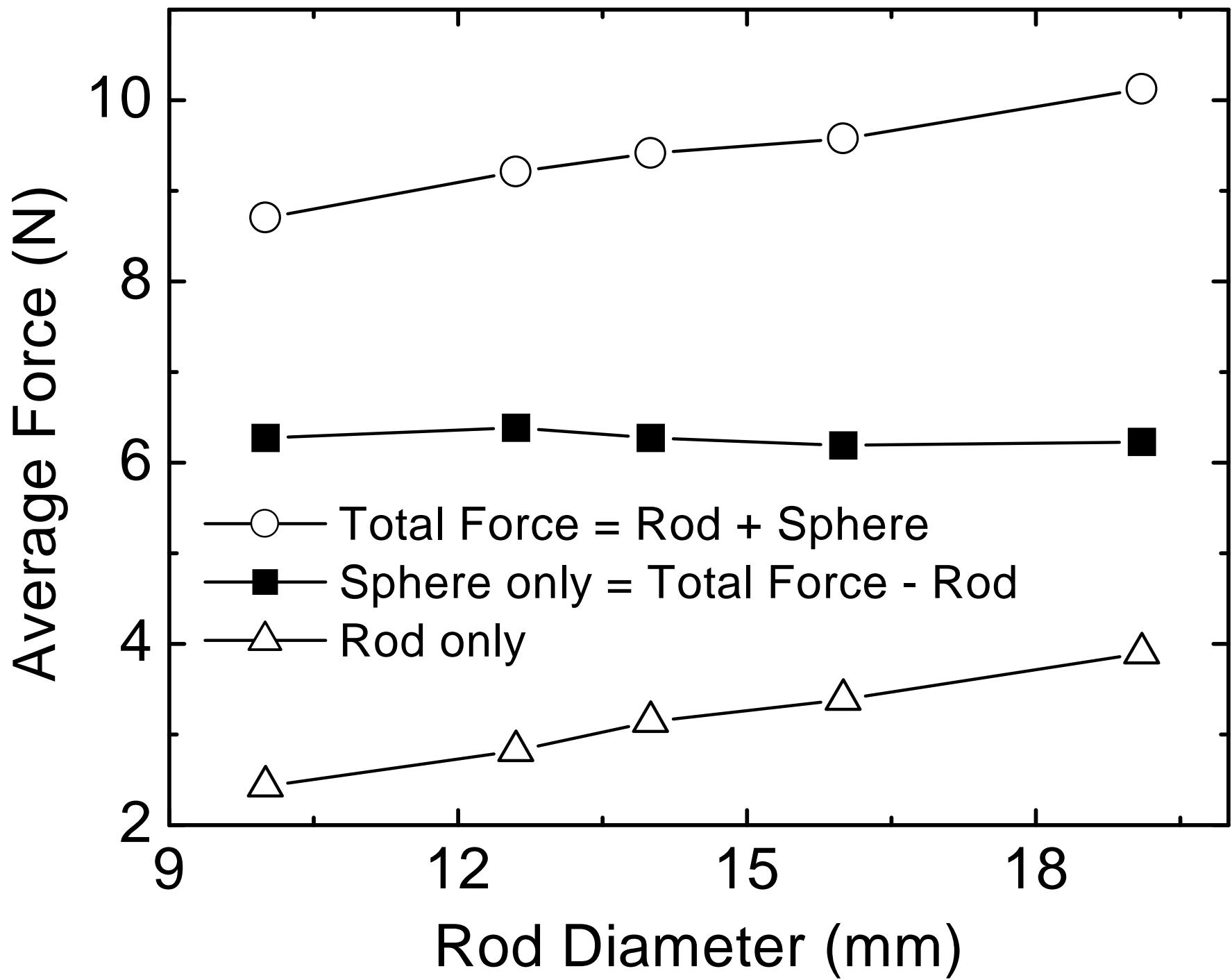
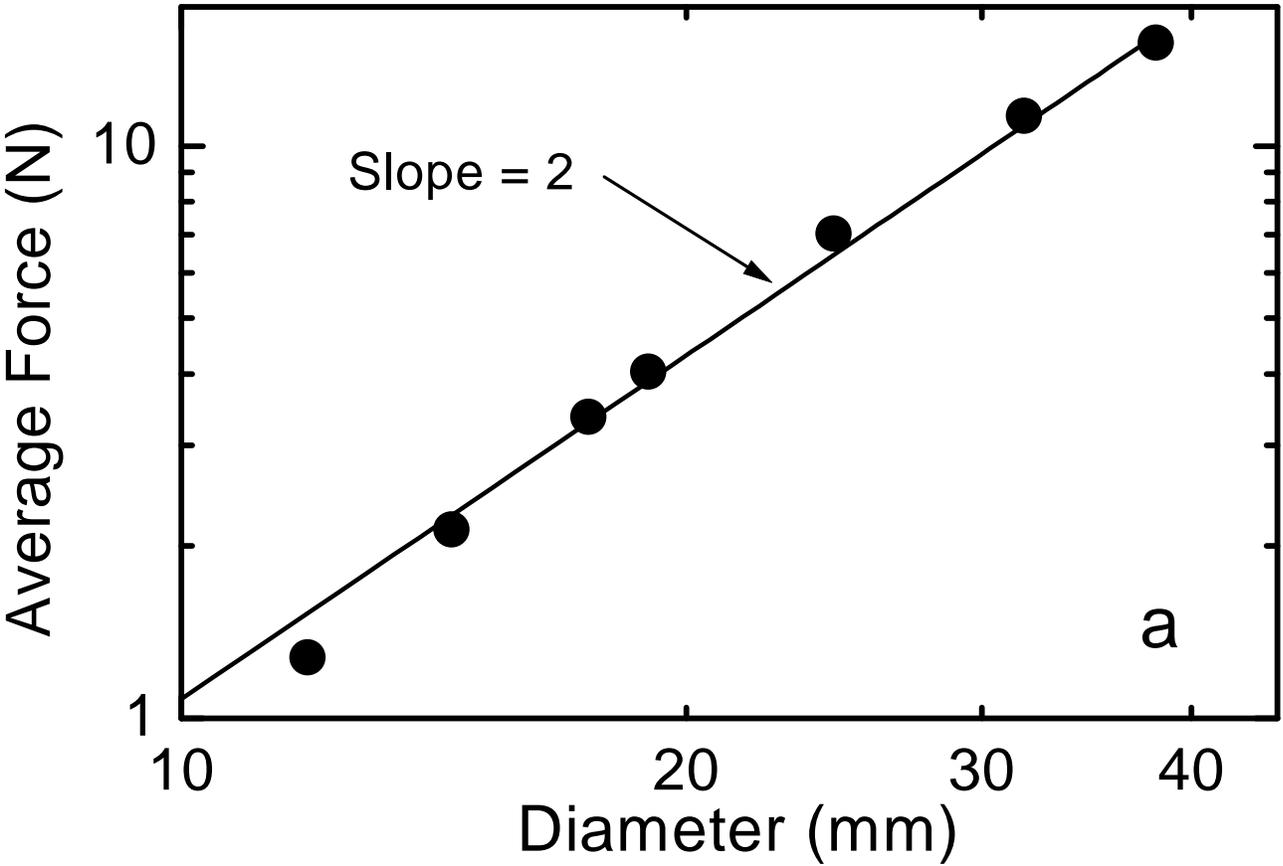
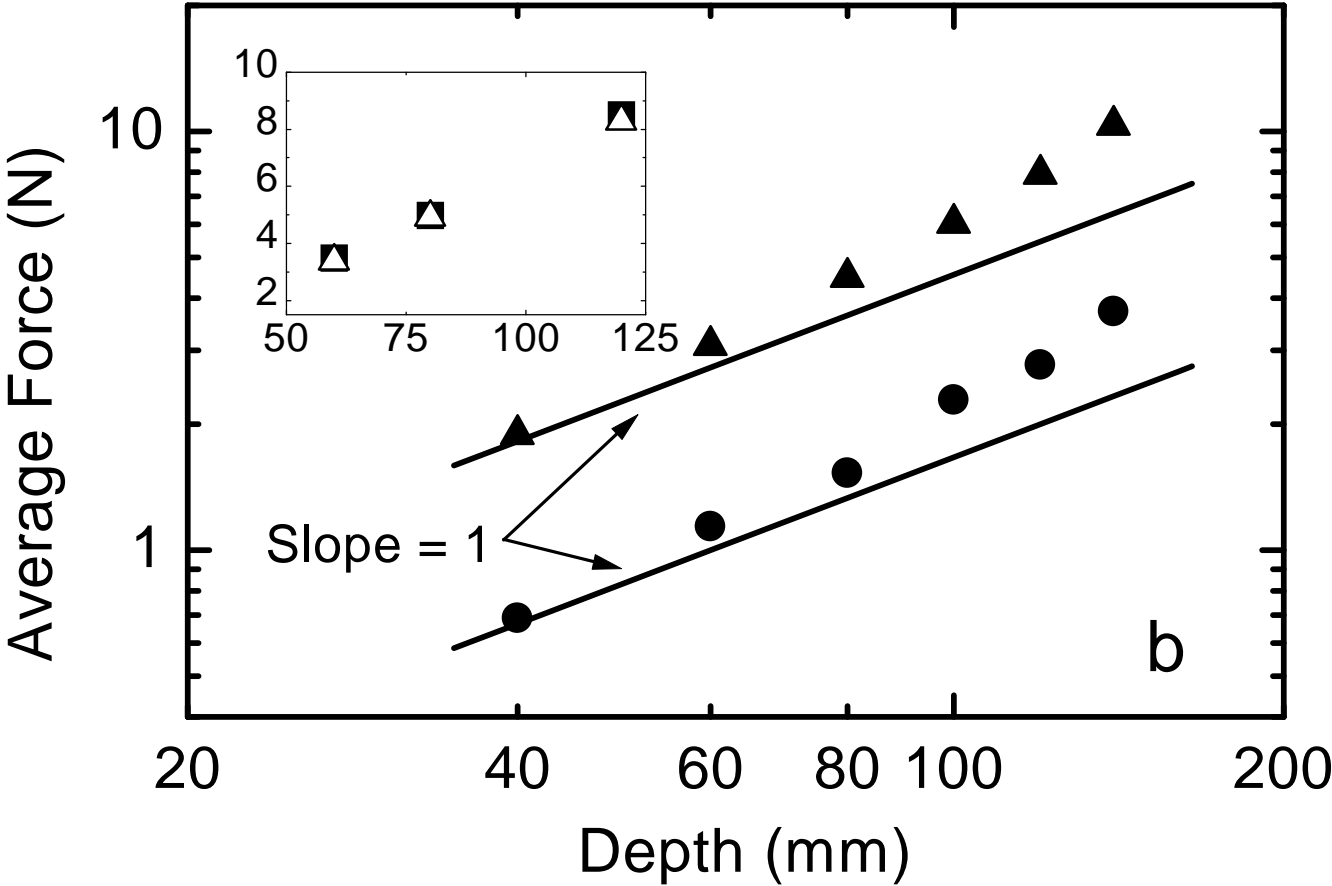
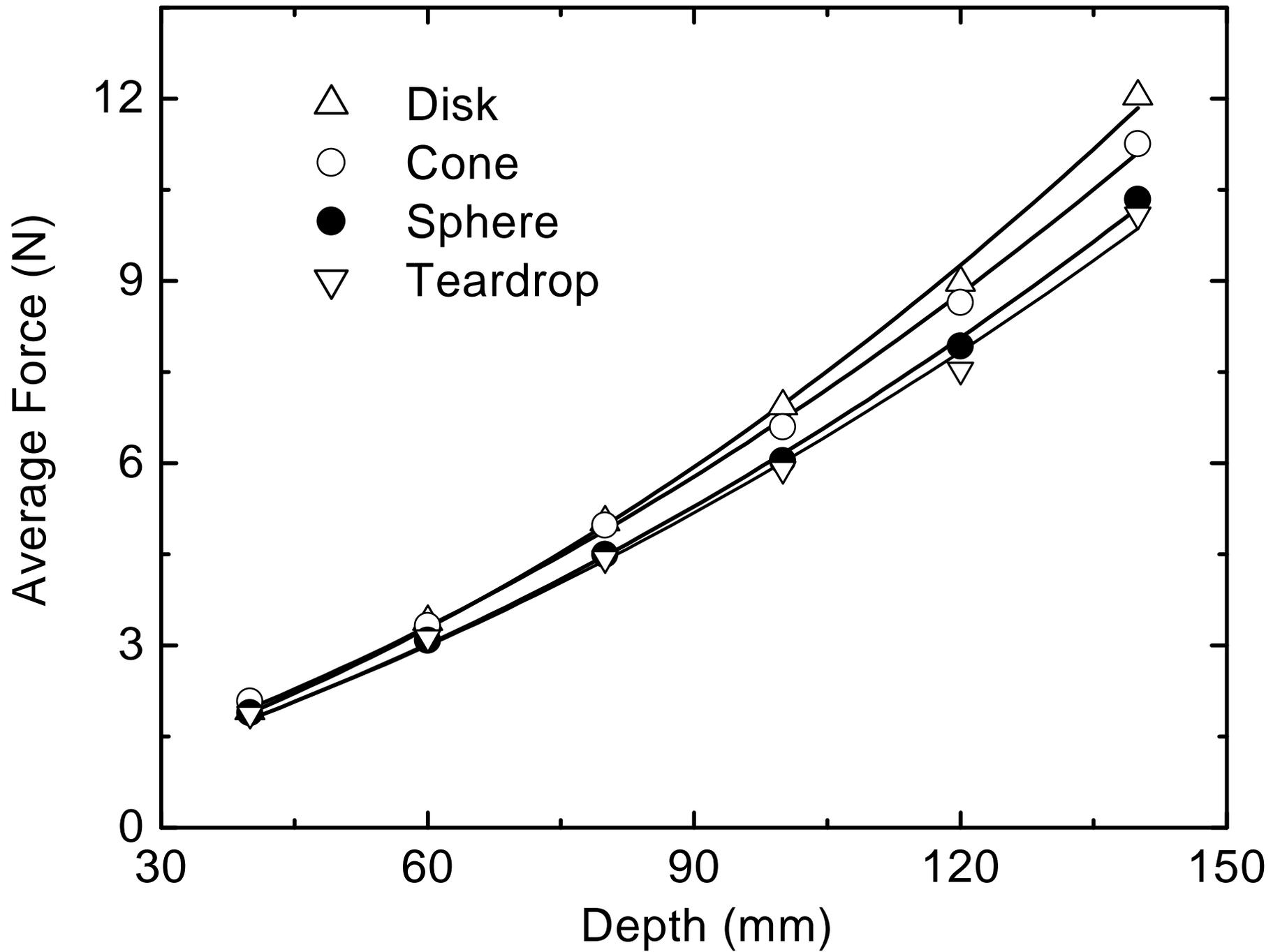
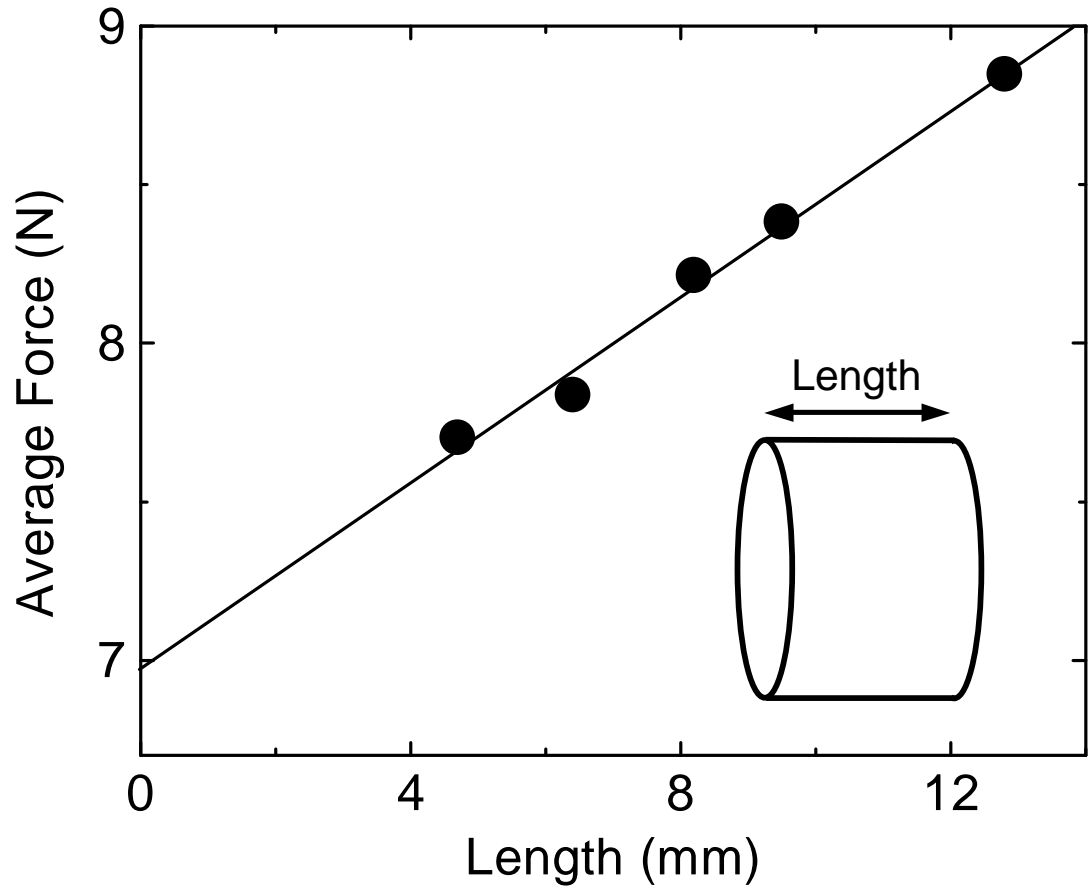
\end{document}